# Web-based 3D and Haptic Interactive Environments for e-Learning, Simulation, and Training

Felix G. Hamza-Lup and Ivan Sopin

Computer Science
Armstrong Atlantic State University

Felix.Hamza-Lup@armstrong.edu

**Abstract.** Knowledge creation occurs in the process of social interaction. As our service-based society is evolving into a knowledge-based society, there is an acute need for more effective collaboration and knowledge-sharing systems to be used by geographically scattered people. We present the use of 3D components and standards, such as Web 3D, in combination with the haptic paradigm, for e-Learning and simulation.

## 1  Introduction

Web-based knowledge transfer is becoming a field of research worthy of attention from the research community, regardless of their domain of expertise, due to the potential of advanced technologies, such as Web 3D and haptics.

In the context of global communication, these technologies are becoming more stimulating, by enabling the creation of collaborative spaces for e-Learning and simulation.

We present several advanced features of Web 3D in conjunction with three successful projects that effectively employ those features. In section 2 we provide a brief introduction to the e-Learning concept. In section 3 we discuss the details of different modalities to enrich user interaction with Web 3D content and haptics. In section 4 we introduce three case studies demonstrating the potential of X3D in simulation and training: 3DRTT, a radiation therapy medical simulator; chemistry and physics concepts interactive simulations project; and HaptEK16, an e-Learning module which provides interaction through haptic feedback for teaching high-school physics concepts. We conclude in section 5 with a set of remarks from our research and development experience.

## 2 Background and Related Work

Let us take a look at the notion of e-Learning. According to [1], the concept of Internet-based learning is broader than Web-based learning, as illustrated in fig. 1. The Web is only one of the Internet services that is based on the HTTP protocol and uses a unified document language, HTML, unified resource locator (URL), and browsers.

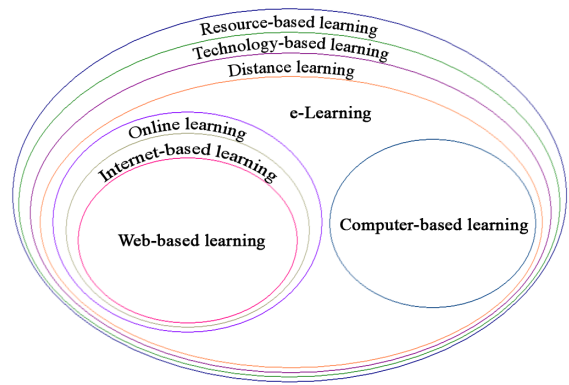

**Fig. 1.** Subsets relationships among the group of terms

As the largest network in the world, the Internet offers several other services besides Web: e-mail, file transfer facilities, etc. Hence, learning can be acquired beyond Web only; correspondence via e-mail makes a good example. Furthermore, the Internet employs a multitude of proprietary protocols along with HTTP.

Due to the advances in 3D technology, it is now possible to develop and deploy on the Web, 3D interfaces and environments that enhance the learning process. An example is Extensible 3D (X3D), an ISO standard for real-time 3D computer graphics and the successor of Virtual Reality Modeling Language (VRML). X3D combines both 3D geometry and runtime behavior descriptions into a single file, with the ability to embed additional scene modules from other sources. X3D files are encoded in either classic VRML or Extensible Markup Language (XML) format. The standard provides means of assigning specific behaviors to 3D objects, enabling users' interaction with them.

On the other hand, advanced interfaces are undergoing a shift towards the incorporation of a new paradigm, haptics. Interfaces combining 3D graphics and haptics have the potential to facilitate our understanding of various concepts and phenomena, as well as, promote new methods for teaching and learning.

Haptic technologies offer new ways of creating and manipulating 3D objects. For instance, in Interactive Molecular Dynamics [2] the users manipulate molecules with real-time force feedback and a 3D graphical display. Another example, SCIRun [3], is a problem-solving environment for scientific computation, which is used to display flow and vector fields, such as fluid flow models for airplane wings.

Initial pilot demonstrations with biology students using augmented graphical models and haptic feedback support the hypothesis that this method provides an intuitive and natural way of understanding difficult concepts and phenomena [4]. Another research group, at the University of Patras, Greece, is involved in designing simulations to aid children comprehend several subject areas of science, such as Newtonian Laws, Space Phenomena, and Mechanics Assembly [5]. Tests show that haptic technology improves the level of human perception due to the deeper immersion provided.

Other fields, such as mathematics and especially geometry, also benefit from haptic interaction. Recently, a system was proposed to allow a haptic 3D-based construction of a geometric problem and its solution representation [6]. Initial performance evaluation indicates the system's elevated user-friendliness and higher efficiency compared to the traditional learning approach.

National Aeronautics Space Administration (NASA) has shown interest in introducing haptics in educational technology. The Learning Technologies Project at the Langley Research Center is concerned with innovative approaches for supporting K-16 education. Pilot study results from the use of simple haptics-augmented machines have yielded positive feedback with 83% of the elementary school and 97% of the college students, rating the software from "Somewhat Effective" to "Very effective" [7] and [8].

## 3  User Interaction

User interaction can be enriched through 3D content and haptics. In what follows we explore in detail the potential of combining X3D with additional Web-enabled instruments, such as HTML and JavaScript, to provide control over a 3D world. We also explain and explore the haptic paradigm and its potential applications in a Web-based environment.

### 3.1  X3D Graphics Visualization

To visualize the graphical content of X3D online, a Web browser needs a special plug-in. Most X3D plug-ins are free for public use and apply a small license fee for commercial use. One example is the BitManagement Contact Player which is both a browser plug-in and a standalone X3D player.

Usually, X3D plug-ins are equipped with a set of basic controls for customizing the user interface and specifying the properties of user interaction: navigational tools, graphics modes, and rendering settings. Being handy, these features only facilitate a user in exploring the visual content, but do not provide any means of altering it. It is the X3D standard itself that allows users to dynamically modify and interact with the 3D graphical scene. There are several alternative ways to implement such systems. In the following subsections we discuss the advantages and drawbacks of a stand-alone X3D-based simulation versus the simulation functionally enriched with JavaScript procedures and HTML versus the haptic-enabled simulation.

**3.2  X3D-based Simulation**

A stand-alone X3D-based environment has all its functionality and graphical user interface (GUI) stored in an X3D file. When a user interacts with the scene, it only responds to the changes specified in the file.

When significant changes in the graphics of the application are required, the code of the file with the graphical content has to be altered. This entails the necessity to provide an updated version of the file and to also manually refresh the X3D scene. Such organization, in general, corresponds to the common client/server interaction on the Web: when an HTML form is populated with data, the user has to press the "Submit" button to request the server's response.

The three-dimensional scope introduced by X3D brings into play new aspects of GUI/user interaction. For instance, volumetric controls, easily implemented in X3D, can better mimic the behavior of the controlled objects. Parts of a 3D world can also be manipulated through a system of specifically designated sensors which respond to clicking, dragging, rotating, and other user actions. The scripting capabilities of X3D can enrich the GUI interactivity as developers have new means to create efficient customized control panels that meet the project-specific tasks.

**3.3  X3D Simulation with HTML and JavaScript Support**

A different approach to improving the GUI interactivity is involving external tools that could effectively communicate with the 3D graphical scene. Good examples of such tools are HTML and JavaScript, most commonly used to build Web pages.

In a GUI with HTML and JavaScript support, JavaScript can be the driving force of most features while HTML only serves as its operating environment. However, HTML makes it difficult to encode unconventional GUI components needed to closer represent the dynamics of the virtual objects in the 3D scene. The HTML standard provides a limited set of traditional input elements, such as checkboxes, buttons, and radio buttons, that are not easily customizable. As a result, the creation of powerful and flexible task-oriented GUI components combines inputs with other traditional HTML objects (layers, images, etc.), backed up with extensive JavaScript code.

With the interface functionality programmed as JavaScript functions, the 3D scene still derives its maneuverability from the methods implemented in the X3D scripting nodes. The browser and X3D environment communicate through mutual function calls. The browser refers to the virtual scene as to an HTML document's object with a number of public functions. Different X3D plug-in developers provide their own sets of such functions. X3D feedback is composed of dynamic injections of JavaScript code. Because both visual content and GUI, are synchronized automatically, no manual page updates are necessary. However, synchronization between HTML and X3D can be an issue in such implementation because it involves continuous calls between the two media and may consume considerable processing power, making webpage rendering slow at the client side.

New client/server communication techniques are also feasible for the development of various dynamic environments. For instance, Asynchronous JavaScript and XML (AJAX) is used in our HTML/JavaScript-based GUI (details in section 4.1) to obtain the listing of external X3D components that could be loaded into the scene. Therefore, without refreshing a page, the user is able to visualize external 3D objects and manipulate them as if they were initially part of the 3D scene.

### 3.4 Multimodal, Haptic-based GUI

Besides traditional GUIs, a novel paradigm, haptic feedback (from the Greek *haptesthai*, meaning "contact" or "touch") may improve the interface usability and interactivity. The tactile sense is the most sophisticated of all our senses as it incorporates pressure, heat, texture, hardness, weight, and the form of objects.

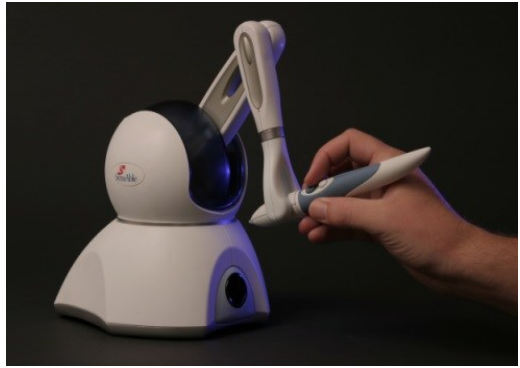

**Fig. 2.** Phantom® Omni™ device

The summarized four basic procedures for haptic exploration, according to [9], each have descriptive object characteristics:

- *Lateral motion* (stroking) provides information about the surface texture of the object.
- *Pressure* gives information about how firm the material is.
- *Contour following* elicits information on the form of the object.
- *Enclosure* reflects the volume of the object.

Current haptic technologies are capable of delivering realistic sensory stimuli at a reasonable cost, opening new opportunities for academic research and commercial developments [10]. Such devices have a distinct set of performance measures [11]:

- *Degrees of freedom* (DOF) are the set of independent displacements that completely specify the position of a body or a system.
- *Workspace* is the volume within which the joints of the device will permit operation.
- *Position resolution* is the minimum detectable change in position within the workspace.

- *Maximum force/torque* is the maximum possible output of the device, determined by such factors as the power of the actuators and efficiency of the gearing systems.
- *Maximum stiffness* is the absolute rigidity of virtual surfaces that can be presented on the device. It depends on the maximum force/torque, but it is also related to the dynamic behavior of the device, sensor resolution, and the sampling period of the controlling processor.

The presence of haptic feedback enables users to feel the virtual objects they manipulate. We have experimented with the PHANTOM® Omni™ device, developed by SensAble Technologies (fig. 2). The Omni™ became one of our choices due to its low cost and force-feedback qualities. It is also backed up by an open source Application Programming Interface (API).

## 4 Case Studies

To illustrate the concepts discussed so far, we will describe three projects developed at our research laboratory (www.cs.armstrong.edu/felix/news) using X3D and haptic technologies: a Web-based medical simulator (3DRTT), chemistry and physics concepts interactive 3D simulations, and a haptic-based module for teaching physics (HaptEK16).

### 4.1 Medical Simulator – 3DRTT

In medicine, success of an operation relies upon practical procedures and the physician's (or surgeon's) experience. Many complex treatment processes are preplanned well in advance of the operation. This is especially the case with radiation therapy. Medical personnel concerned with the planning part (e.g., correct radiation dosages and appropriate patient setup) are sometimes frustrated by the fact that a theoretically sound plan proves inconsistent with the current hardware and patient constraints (i.e., collisions with the patient and treatment hardware may occur). This issue can be addressed by complementing clinical setups with effective visual simulations.

3D Radiation Therapy Training (3DRTT) is a Web-based 3D graphical simulator for radiation therapy procedures. It simulates linear accelerators (linacs) used to operate patients that have cancer by delivering radiation doses to an internal tumor. The project focuses on improving the efficiency and reliability of the radiation treatment planning and delivery by providing accurate visualization of the linacs hardware components, as well as, careful imaging of their interactive motion.

The virtual representation of the treatment settings (fig. 3) provides patients and therapists with a clear understanding of the procedure. Equipped with the patient Computer Aided Tomography (CAT) scan data, the treatment planner can simulate a series of patient-specific setups and detect unforeseen collision scenarios for complex beam arrangements. Hence, the necessary adjustments in the treatment plan can be made beforehand and validated.

Another important application of 3DRTT is improving the current level of radiation therapy education, training, and safety. With such a Web-based 3D simulation tool at the disposal of the radiation therapy staff, there is plenty of room for exploring various treatment procedures (linac components, motion limitations, associated accessories, etc.) and gaining experience for future operations. Moreover, exposure of trainees to harmful radiation and accidents are avoided.

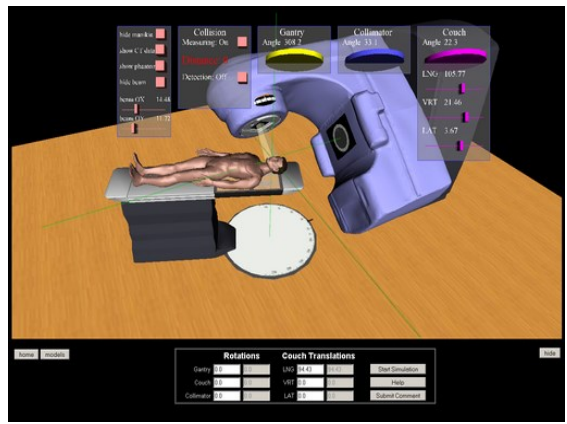

**Fig. 3.** 3DRTT simulator with X3D-based GUI.

Currently, two versions of the simulator, with X3D and HTML/JavaScript-based GUIs (refer to sections 3.2 and 3.3), are available on the project's website (www.3drtt.org). The X3D-based version provides tools for controlling the angles and locations of the parts of the machine.. The GUI is composed of several semitransparent panels containing various volumetric controls. The controls are designed to logically correspond to the assigned operations (specifically, scrolls for rotations, sliders for translations, and buttons for switching among different simulation modes) and therefore improve the interface usability [12]. GUI components can be rearranged dynamically to avoid occlusions of important parts of the scene.

The HTML/JavaScript-based GUI alternative supports the same functionality and provides additional features (fig. 4). Instead of floating X3D menus, the simulator controls are shifted to the scope of the HTML page. The set of GUI elements includes sliders, buttons, and displays that enhance the interface learnability. For the convenience of navigation in the virtual space, the control panel can be hidden and brought back at the user's request.

Current Web technologies, such as AJAX, introduce new methods of accessing and dynamically processing external modules. For instance, the user may load various hardware attachments for the linacs directly in the virtual world. The source X3D file for the simulator does not "know" how many attachments are available at the moment and what their names are. However, upon user's request, an AJAX function makes a call to a scriptlet stored on the server and receives the listing of available files as the response. This listing is transmitted to an X3D script that handles the loading and embedding of the specified files into the virtual environment. Therefore, no alterations of the X3D source code are necessary when new attachments are uploaded to the server because they become immediately accessible. This dynamic behavior enhances the interactivity of simulation as users can have access to external modules without the need to manually search for them or modify the scene configuration.

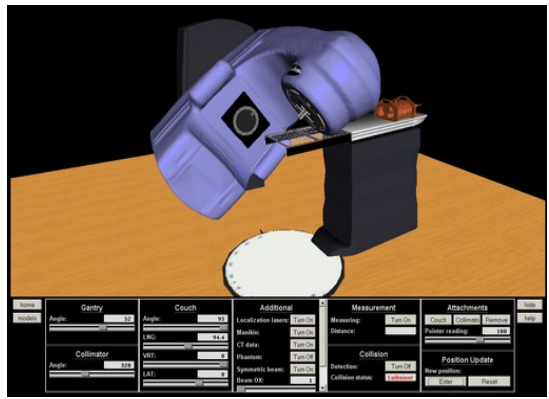

**Fig. 4.** 3DRTT Simulator with HTML/JavaScript-based GUI.

### 4.2   Chemistry and Physics Concepts Interactive 3D Simulations

The first module of this project is a virtual demonstration of the electrolysis concept. Electrolysis is the technique of separating compound chemical elements into simpler elements with opposite charges by applying an electric current. Our simulation offers an intuitive and effective visualization of the process by replicating the behavior of the decomposed molecules. The interface (fig. 5) includes a 3D tank model with experimental solution ($H_2O$ or NaCl), a cathode and anode dipped into the solution, a cord connecting the cathode and anode to a bulb, and colored spheres representing the atoms of sodium and chlorine. There is also a floating menu with a legend explaining the appearance of the atoms, molecules, and ions at different stages, and a slider to control the speed of simulation.

  The graphical scene is embedded into an HTML page which contains *Start* and *Reset* buttons and provides a textual explanation of the phenomenon. Once the simulation begins, the molecules of NaCl break into positively charged ions of sodium and negatively charged ions of chlorine:

$$NaCl = Na^+ + Cl^- .  \quad\quad\quad\quad\quad (1)$$

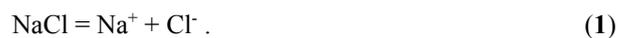

Then ions of sodium are attracted to the cathode where they lose an electron and become neutral atoms; and ions of chlorine are attracted to the anode where they obtain an electron, become neutrally charged atoms which form molecules of $Cl_2$ and evaporate based on the formula:

$$2NaCl \rightarrow 2Na + Cl_2\uparrow .  \qquad (2)$$

All steps of the reaction are explicitly reflected in the simulation by the motion of reagents' atoms. As an additional visual cue, the bulb radiates small portions of light, indicating the presence of electric current.

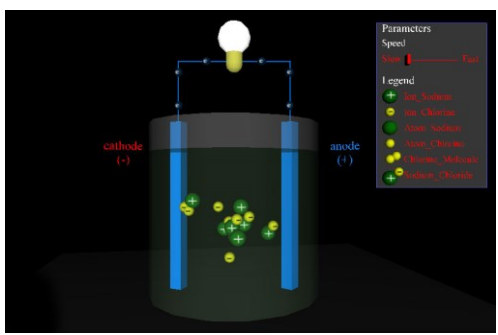

**Fig. 5.** NaCl electrolysis simulation

The simulation demonstrates the effectiveness of a Web 3D interface employed as a part of Web page multi-media content. In addition, animated X3D scene introduces a fair level of abstraction that helps convey the nature of the phenomenon without confusing the user with redundant graphical complexity.

### 4.3 Haptic e-Learning – HaptEK16

HaptEK16 [13] is designed to assist students in understanding Pascal's principle and other difficult concepts of hydraulics. The simulator includes three modules: pressure measurement, hydraulic machine, and hydraulic lifting simulation. Students can interact with the 3D scene using a haptic device, as illustrated in fig. 6. The functionality of the simulator is implemented through Python, X3D, and the Sense Graphics' H3D API (www.sensegraphics.com), discussed further.

Python is an object-oriented scripting language that offers strong support for integration with other toolkits and APIs. According to InfoWorld [14], Python's user base nearly doubled in 2004 and currently includes about 14% of all programmers. Python is available for most operating systems, including Windows, UNIX, Linux, and Mac.

Some of the Python's strengths which were considered when selecting a language to implement the system functionality include:

- *Low complexity*: wxPython (an auxiliary library for GUI) was selected because of its ease of use and reduced complexity compared with Java/Swing;
- *Prototyping*: Prototyping in Python is quick and simple and often leads to a quick prototype that can be adapted for the development of the final system;
- *Maintainability*: The code in Python is easy to modify and/or refactor. Less time is spent understanding and rewriting code which leads to an efficient integration of new features.

H3D is an open X3D-based haptic API. It is written entirely in C++ and uses OpenGL for graphics rendering and OpenHaptics (de-facto industry standard haptic library) for haptic rendering. With its haptic extensions to X3D, the H3D API is an excellent tool for writing haptic-visual applications that combine the sense of touch and 3D graphics. The main advantage of H3D to OpenHaptics users is that, being a unified scene graph API, it facilitates the management of both graphics and haptics rendering.

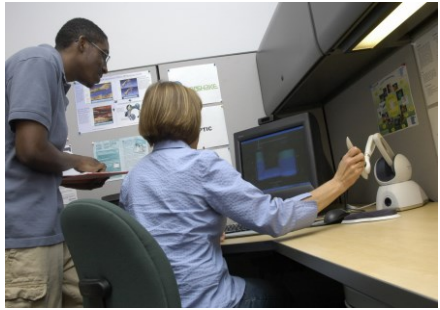

**Fig. 6.** Students using the HaptEK16 hydraulics demo

The scene graph concept facilitates application development, but it can still be time consuming. SenseGraphics extended their API with scripting capabilities in order to empower the user with the ability of rapid prototyping. The design approach used in HaptEK16 was the one recommended by SenseGraphics; i.e., geometry and scene-graph structure for a particular application were defined using X3D, and application and user interface behaviors were described using Python and wxPython.

Programming the sense of touch for a virtual object involves two steps. First, the programmer must specify the haptic device to use; second, a set of haptic properties must be defined for each "touchable" object. To specify the haptic device, an instance of the *DeviceInfo* node is created, and the haptic device is added to it. *HLHapticsDevice* is the node used to manipulate a Phantom device. The graphical representation of the device (in case of HaptEK16, a sphere) is also specified in the *containerField* group, as illustrated next.

Example of specifying the haptic device in an X3D file

```
<DeviceInfo>
  <HLHapticsDevice positionCalibration="
      1e-3 0 0 -.15
      0 2e-3 0 .05
      0 0 1e-3 0
      0 0 0 1">
    <Group containerField="stylus">
      <Shape>
        <Appearance>
          <Material />
        </Appearance>
        <Sphere radius="0.0025" />
      </Shape>
      <Transform translation="0 0 0.08"
                 rotation="1 0 0 1.570796">
        <Shape>
          <Appearance>
            <Material />
          </Appearance>
          <Cylinder  radius="0.005"
                     height="0.1" />
        </Shape>
      </Transform>
    </Group>
  </HLHapticsDevice>
</DeviceInfo>
```

To implement the tactile sensation for a generic shape, one must add a surface node with haptic properties to the shape's *Appearance* node. In HaptEK16 this is accomplished with a frictional surface node added to the cylinder's *Appearance* node. The *DynamicTransform* node is added to define properties for rigid body motion.

Example of implementing haptic properties in an X3D file

```
<DynamicTransform DEF="DYN1"
  mass=".05"
  inertiaTensor=".1 0 0 .1 0 0 0 .1">
  <Shape>
    <Appearance>
      <Material diffuseColor="0 .8 .8" />
      <FrictionalSurface dynamicFriction=".6"
                         staticFriction=".2" />
    </Appearance>
    <Cylinder DEF="LEFTCYL"  height=".085"
                             radius=".045" />
  </Shape>
</DynamicTransform>
```

The X3D file format is used by H3D as an easy way to define geometry and arrange scene-graph elements, such as user interfaces. A screenshot of the HaptEK16 e-Learning module is illustrated in fig. 7.

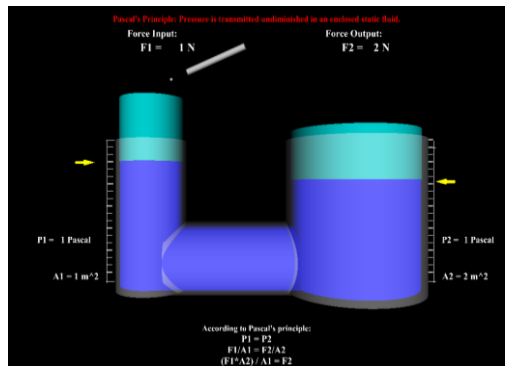

**Fig. 7.** HaptEK16 screenshot and corresponding Phantom® Omni™ Device

A set of test questionnaires were designed and implemented for the assessment of the e-Learning module. The results from the assessment tests proved that the student group exposed to the HaptEK16 simulator scored better (13% higher total scores) than the group that was not. Such results indicate the potential of using X3D and haptics to develop novel simulation and training environments.

## 5   Conclusions

3DRTT serves as an example of a Web-based system extensively taking advantage of X3D to improve the efficiency of the user-interface interaction, as well as, to provide powerful means of professional education and training. Naturally, complex concepts and settings are better understood when delivered with visual support, especially in complicated scenarios. Easy online access, simple control, and advanced capabilities of 3D visualization of radiation therapy treatment scenarios proved to be of great value to the radiation therapists using our system. Currently, 3DRTT has over a hundred registered users and keeps attracting the attention of other professionals working in the radiation therapy field.

Another project, chemistry and physics concepts interactive 3D simulations, proves the effectiveness of X3D interactive models embedded in a Web page content to support the description of the presented phenomenon. X3D scene visually depicts the concept that the user reads about on the very same page. Therefore, Web 3D significantly deepens the level of learner's comprehension.

The other development, the haptic e-Learning module (HapteK16) facilitates student understanding of difficult concepts (e.g., in science) and has the potential to augment or replace traditional laboratory instruction with an interactive and cost-effective interface offering enhanced motivation, retention, and intellectual stimulation. HaptEK16's haptics-augmented activities allow students to interact and feel the effects of forces in the experiment. We believe that force feedback will lead to more effective learning and that the HaptEK16 project and similar projects have a substantial and still unexplored educational potential.

Considering the advances in software and hardware technology, we foresee many applications of haptics and 3D graphics in the near future, broadening the communication channels among people and narrowing the knowledge gap among us.